# Extremely-Fast, Energy-Efficient Massive MIMO Precoding with Analog RRAM Matrix Computing

Pushen Zuo, Zhong Sun, *Member, IEEE*, and Ru Huang, *Fellow, IEEE*

*Abstract*—Signal processing in wireless communications, such as precoding, detection, and channel estimation, are basically about solving inverse matrix problems, which, however, are slow and inefficient in conventional digital computers, thus requiring a radical paradigm shift to achieve fast, real-time solutions. Here, for the first time, we apply the emerging analog matrix computing (AMC) to the linear precoding of massive MIMO. The real-valued AMC concept is extended to process complex-valued signals. In order to adapt the MIMO channel models to RRAM conductance mapping, a new matrix inversion circuit is developed. In addition, fully analog dataflow and optimized operational amplifiers are designed to support AMC precoding implementation. Simulation results show that the zero-forcing precoding is solved within 20 ns for a 16×128 MIMO system, which is two orders of magnitude faster than the conventional digital approach. Meanwhile, the energy efficiency is improved by 50×.

*Index Terms*—Massive MIMO, precoding, analog matrix computing, RRAM, memristor.

## I. INTRODUCTION

MASSIVE multiple input multiple output (MIMO) is an enabling technology of the $5^{th}$ generation (5G) mobile network. Compared with the conventional MIMO used in the earlier generations, such as 4×4 or 8×8 antenna configurations in LTE-advanced 4G, the number of antennas in massive MIMO is radically increased [1]. As a result, the spectral capacity, data rate, and energy efficiency of wireless communications are significantly improved [2], [3]. Looking forward to the upcoming 6G, the ultra-massive MIMO is expected to be equipped with >1000 antennas for further performance improvements [4].

To deliver the desired throughput and efficiency of massive MIMO, precoding is usually used to minimize the interference between individual user channels. Compared with the complicated, expensive dirty-paper coding method, zero-forcing (ZF) linear precoding is advantageous in terms of computing resource demands, while achieving near optimal performance [5]. ZF precoding is essentially about solving a system of linear equations (or a generalized inverse matrix problem) in the complex number domain, which features a cubic computational complexity and thus imposing a heavy workload to the baseband processors. To accelerate ZF precoding in the conventional digital paradigm, various algorithmic optimizations have been proposed, such as QR decomposition [6], Gauss-Jordan elimination [7], and Neumann series [8]-[10]. However, the optimization is fundamentally limited by the binary representation, logic gates, and sequential processing, which prevent the reduction of the order of digital algorithm complexity. In addition, digital computers adopt the von Neumann architecture, where the communications between the memory and the processor cause extra costs of time and energy that are aggravated in data-intensive applications [11].

Recently, analog matrix computing (AMC) with resistive memory devices has been developed for fast, efficient matrix computations [12]. A device prototype is the resistive random-access memory (RRAM), which features simple device structure and good analog conductance capability. A crosspoint RRAM array represent naturally a physical analog matrix, based on which the basic matrix operations can be realized, including matrix multiplication [13], matrix inversion and eigenvector [14], and generalized inverse [15]. Due to the high spatial parallelism of the crosspoint array, the time complexity of AMC is dramatically reduced, literally reaching $O(1)$ [16]-[18]. As the computation is carried out *in situ* in the memory array (and the peripheral circuits), this approach realizes in-memory computing to save communication costs.

In this work, for the first time, we apply the AMC circuits to the precoding of massive MIMO. We designed an AMC-based linear precoding architecture, by developing a fully analog dataflow including the matrix multiplication and inversion circuits. The complex-valued matrix/vector data can be easily accommodated in the AMC circuits for fast computations. According to the parameter distribution of the MIMO channel model, the matrix inversion circuit is modified to take full advantage of the RRAM conductance range. Based on the RRAM model and the optimized operational amplifiers, ZF precoding with the AMC architecture is completed within only 20 ns, demonstrating significant advantages of computing speed and energy efficiency over traditional digital approaches. The bit error rate is also analyzed to validate the adequacy of AMC for reliable wireless communications.

## II. SYSTEM DESCRIPTION

### A. Channel model

In a massive MIMO system (Fig. 1), a channel matrix $\boldsymbol{H} \in \mathbb{C}^{K \times M}$ is used to describe the channel gains between the base station (BS) and the user equipments (UEs), where $M$ is the number of BS antennas, $K$ is the number of UEs (Fig. 1a). Usually, $M$ is much larger than $K$. An entry $h_{k,m}$ represents the channel gain between the $m$-th BS antenna and the $k$-th UE. $\boldsymbol{x} =$

This work was supported in part by the National Key R&D Program of China under Grant 2020YFB2206001, in part by NSFC under Grant 62004002, Grant 92064004 and Grant 61927901, and in part by the 111 Project under Grant B18001.

Authors are with the Institute for Artificial Intelligence, the School of Integrated Circuits, Peking University, Beijing Advanced Innovation Center for Integrated Circuits, Beijing 100871, China, (e-mail: zhong.sun@pku.edu.cn).

$[x_1, x_2, ..., x_M]^T$ is the transmission data of $M$ BS antennas, and the symbol signals received by the $K$ UEs is

$$\mathbf{y} = \sqrt{\rho_T}\mathbf{Hx} + \mathbf{n}, \quad (1)$$

where $\rho_T$ is the transmission power, $\mathbf{n} \sim \mathcal{CN}(0, \sigma^2 \mathbf{U}_K)$ is the additive white Gaussian noise with $\mathbf{U}_K$ being the $K \times K$ identity matrix. The signal-noise ratio (SNR) in dB is defined as $SNR = 10\log_{10}\frac{MP_{tx}}{\sigma^2}$. Before the transmission of vector $\mathbf{x}$, the precoding of symbols is completed in the BS, that is

$$\mathbf{x} = \mathbf{Ws}, \quad (2)$$

where $\mathbf{s}^{K \times 1}$ is the symbol vector destined to the UEs, and $\mathbf{W}^{M \times K}$ is the precoding matrix. For ZF precoding, there is

$$\mathbf{W} = \frac{1}{\alpha}\mathbf{H}^H(\mathbf{HH}^H)^{-1}, \quad (3)$$

where $\alpha$ is a precoding factor for satisfying the power constraint, $(\cdot)^H$ and $(\cdot)^{-1}$ denote Hermitian transpose and matrix inversion, respectively.

In reality, $\mathbf{H}$ is obtained through channel estimation carried out in the BS, although simplified models are usually used for theoretical investigations. Here, we consider the Rayleigh fading channel model, where entries of matrix $\mathbf{H}$ are independent identically distributed (i.i.d.) Gaussian random variables, i.e., $h_{k,m} \sim \mathcal{CN}(0,1)$. Symbol vector $\mathbf{s}$ is the result of quadrature amplitude modulation (QAM). For the 16-QAM employed in this work, each element $s_i$ is given by

$$s_i = \beta(r_i + jt_i), i = 1,2,...,K, \quad (4)$$

where $r_i, t_i \in \{\pm 1, \pm 3\}$, and $\beta$ is the normalization coefficient.

### B. AMC implementation

The expression of Eq. (3) is actually the generalized right inverse of matrix $\mathbf{H}$. Despite an AMC circuit is capable of this computation [12], its implementation requires mapping two copies of $\mathbf{H}$, which in turn contains positive/negative real and imaginary parts, resulting in a high-complexity circuit with complicated connections. Instead, we consider to realize Eq. (3) in two steps, including one matrix inversion and one matrix multiplication, which both can be realized in the AMC paradigm. Specifically, the first step is to obtain the intermediate result $\mathbf{y} = \mathbf{Z}^{-1}\mathbf{s}$ through the matrix inversion (INV), where $\mathbf{Z} = \mathbf{HH}^H$ is a Gram matrix. The second step is to calculate $\mathbf{x} = \mathbf{H}^H\mathbf{y}$ through the matrix-vector multiplication (MVM). The proposed AMC-based precoder is shown in Fig. 1, mainly consisting of an INV circuit, an MVM circuit, and a sample & hold (S&H) module for analog voltages transmission.

### III. CIRCUIT DESIGN AND ANALOG MAPPING

Matrices of MIMO model are complex-valued, described as $\mathbf{H} = \mathbf{R}_H + j\mathbf{I}_H$ and $\mathbf{Z} = \mathbf{R}_Z + j\mathbf{I}_Z$. To enable the complex matrix computation in the real-number domain, real-valued matrices should be constituted, namely $\mathbf{\Omega}_H = \begin{bmatrix} \mathbf{R}_H & -\mathbf{I}_H \\ \mathbf{I}_H & \mathbf{R}_H \end{bmatrix}$ and $\mathbf{\Omega}_Z = \begin{bmatrix} \mathbf{R}_Z & -\mathbf{I}_Z \\ \mathbf{I}_Z & \mathbf{R}_Z \end{bmatrix}$. Correspondingly, vectors $\mathbf{s}$, $\mathbf{y}$, $\mathbf{x}$ should also be expanded, for instance, $\mathbf{\Omega}_s = \begin{bmatrix} \mathbf{R}_s \\ \mathbf{I}_s \end{bmatrix}$.

### A. INV and MVM circuits

According to the definition of Rayleigh fading model, the distribution of matrix $\mathbf{H}$ entries follows the Gaussian distribution. However, due to the special structure of the Gram matrix $\mathbf{Z}$, the distribution of matrix $\mathbf{Z}$ is non-trivial. Take the case of $K=16$, $M=128$ as an example, the resulting matrix $\mathbf{Z}$ is shown in Fig. 2a. $\mathbf{Z}$ appears to be diagonally dominant, which is an important feature of massive MIMO and eventually facilitates the fast response of the INV circuit. By running many random tests, the distribution of matrix $\mathbf{Z}$ is obtained (Fig. 2b), which typically shows two bell curves with a separation range.

To store the matrix $\mathbf{Z}$ in an RRAM array, it is appropriately scaled. In this case, $\mathbf{Z}$ is downscaled by 1/64, which renders the first bell curve collapsing within ±0.5 faithfully. Meanwhile, the second bell curve becomes centered around 2. Due to the discontinuous distribution of $\mathbf{Z}$ entries, direct mapping to RRAM device conductance is inefficient. To solve this issue, we consider shifting the second bell curve to be zero-centered by subtracting a constant, resulting in the distribution in the inset of Fig. 2b. Furthermore, it is observed that the larger values are all contributed by the diagonal entries. Given that $\mathbf{Z}$ is a Gram matrix, the diagonal entries are all positive real numbers. Consequently, shifting the curve is equivalent to subtracting a constant diagonal matrix from matrix $\mathbf{\Omega}_Z$, resulting in $\mathbf{M} = \mathbf{\Omega}_Z - \mathbf{D}$, where $\mathbf{D} = 2\mathbf{U}_{2K}$ is the identity matrix times 2. Then, to map matrix $\mathbf{M}$ in crosspoint RRAM arrays, it is split conventionally by two non-negative matrices, i.e., $\mathbf{M} = \mathbf{A}_{INV} - \mathbf{B}_{INV} = \begin{pmatrix} \mathbf{R}_M^+ & \mathbf{I}_M^- \\ \mathbf{I}_M^+ & \mathbf{R}_M^+ \end{pmatrix} - \begin{pmatrix} \mathbf{R}_M^- & \mathbf{I}_M^+ \\ \mathbf{I}_M^- & \mathbf{R}_M^- \end{pmatrix}$, where $\mathbf{R}_M^+$ ($\mathbf{R}_M^-$), $\mathbf{I}_M^+$ ($\mathbf{I}_M^-$) are the absolute positive (negative) real part, positive (negative) imaginary part of $\mathbf{M}$, respectively.

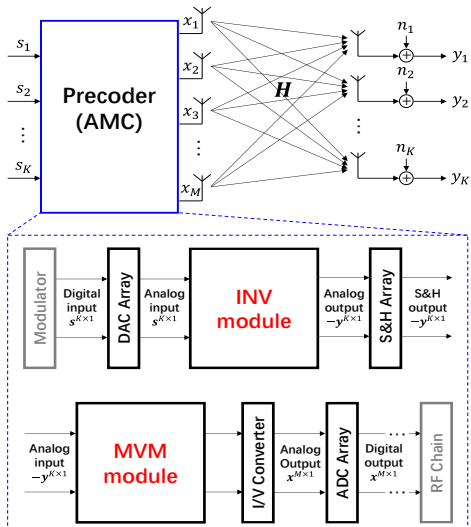

Fig. 1. AMC-based massive MIMO precoder.

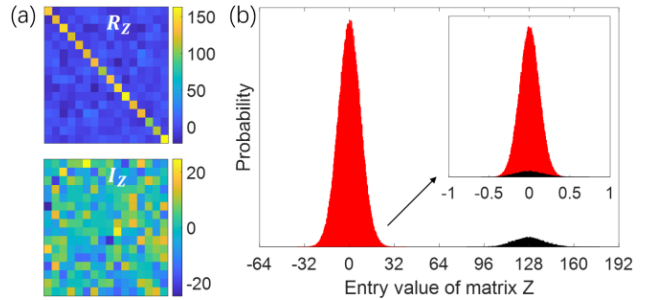

Fig. 2. (a) The real part and the imaginary part of a Gram matrix $\mathbf{Z}$ of the Rayleigh fading model. $K=16$, $M=128$. (b) Distribution of entry values of matrix $\mathbf{Z}$ (results from 1000 numerical tests).

Based on the shifting and splitting scheme, a modified INV circuit is proposed to suit the matrix inversion in precoding (Fig. 3a). Two crosspoint RRAM arrays are used to map $A_{INV}$ and $B_{INV}$, respectively, the subtraction between which two is realized by a row of analog inverters. Additionally, a resistor array is used to map the diagonal matrix $D$. The input vector $\Omega_s$ is provided to crosspoint rows as injected currents. A column of operational amplifiers (OAs) connect the crosspoint rows and columns to form feedback loops, thus enabling the fast solution to the intermediate result $\Omega_y$. Due to the virtual short of OAs, the current summations at the crosspoint rows must satisfy $(A_{INV} + D - B_{INV}) \cdot (-\Omega_y) + \Omega_s = 0$. Therefore, the output voltages of OAs represent the real and imaginary part of the solution $y = Z^{-1}s$. Since the Gram matrix $Z$ is positive definite axiomatically, the stability of the INV circuit is guaranteed.

In the circuit, the input vector $s$ should be scaled to align with the matrix scaling. In the specific case, $s$ is downscaled by 1/8, to facilitate the medium output voltages that take full advantage of the available range while prevent the RRAM devices from disturbance. As a result, the output vector $y$ obtained in the circuit is actually downscaled by 1/8. We observed that matrix $Z$ entry values increase with the number $M$, which in turn suggests a proportionally different scaling factor. Despite a specific MIMO channel model is targeted in this work, many other common models such as correlation channel model, Kronecker model and Weichsel Berger model [19] show similar matrix distributions, which is fundamentally contributed by the Gram matrix nature of $Z$.

The output vector $\Omega_y$ of the INV circuit shall be multiplied by matrix $\Omega_H$. To do so, $\Omega_y$ needs to be transferred as the input vector of the MVM circuit. To avoid the large-area and energy-hungry analog/digital converters accompanying each matrix operation, we use analog S&H circuits to seamlessly transfer the intermediate results (Fig. 3b) [20]. The received vector $\Omega_y$ and its opposite are applied to two crosspoint RRAM arrays of $\Omega_H$ (Fig. 3c). The matrix $\Omega_H$ is represented by two the difference between two non-negative matrices, namely $\Omega_H = A_{MVM} - B_{MVM} = \begin{pmatrix} R_H^+ & I_H^- \\ I_H^+ & R_H^+ \end{pmatrix} - \begin{pmatrix} R_H^- & I_H^+ \\ I_H^- & R_H^- \end{pmatrix}$, where $R_H^+$, $R_H^-$, $I_H^+$, and $I_H^-$ are similar to the case of matrix $Z$. Since MVM is performed with the Hermitian transpose of $H$, namely $H^H$, the input vector is conveniently applied to the rows, and the outputs are collected at the columns. In the analog MVM, the matrix and vector should be scaled as well. In this case, $H$ is downscaled by 1/8, and the input vector $y$ conveyed from the INV output is already downscaled by 1/8. As a result, the output of MVM circuit represents the correct result $x$ without any scaling.

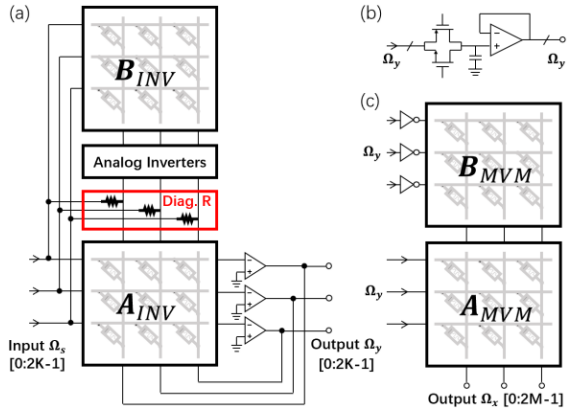

Fig. 3. (a) Matrix inversion circuit for obtaining the intermediate vector $y^{K \times 1}$. The resistors marked by the red box map the diagonal matrix $D$. (b) S&H circuit for transferring vector $y$, with a transmission gate for control. (c) MVM circuit for computing solution $x$.

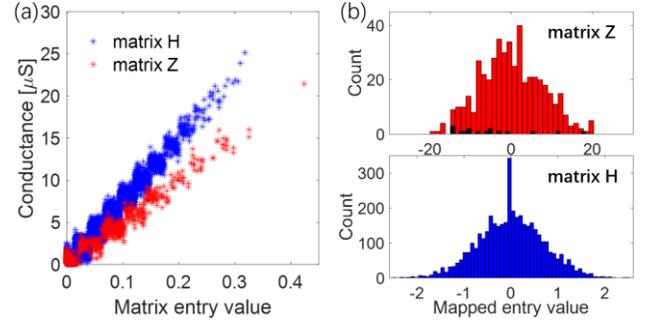

Fig. 4. (a) Results of mapping matrices $A_{INV}$ and $B_{INV}$ (of $Z$), $A_{MVM}$ and $B_{MVM}$ (of $H$) to RRAM conductances. (b) Distributions of the mapped matrices $Z$ (after removing the diagonal matrix) and $H$.

### B. RRAM model and OA design

To map the real-valued matrices $\Omega_Z$ and $\Omega_H$, we consider the RRAM conductance range of 0.1-30 μS, which is typical of many RRAM device candidates [21]. 15 uniform discrete states across 2-30 μS and 1 deep high resistance state are assumed to model a 4-bit RRAM. Each level follows a normal distribution to account for the programming error, device noise, and conductance drift. To fully exploit the conductance range, an appropriate reference conductance value should be selected for mapping $\Omega_Z$ in INV or $\Omega_H$ in MVM. Representative mapping results are shown in Fig. 4a. Based on the mapping results, matrices $Z$ and $H$ are recovered in Fig. 4b, which are in line with the ideal distributions.

In parallel with the crosspoint RRAM arrays, OA is another basic module in AMC, since it is involved in all the circuits in Fig. 3. To increase the open-loop gain, the 2-stage folded cascode structure is used for OA design (Fig. 5). Before the class-A output stage, a Miller capacitor is connected to improve the slew rate, which is beneficial to the response speed of the circuits. The supply voltage of OA is ± 0.6 V. Simulation results show that the open-loop gain is 50.5 dB, and the unity gain bandwidth is 157 MHz. By adjusting the aspect ratios of transistors in the output stage of OA, OA's load capacity can be adjusted to meet the different requirements for the output stage currents in INV and MVM circuits. The OAs in the S&H module adopt the rail-to-rail structure, by adding a differential pair as marked in Fig. 5, which increases the common-mode input voltage range and the output voltage swing. Simulation results of rail-to-rail OA show that the open-loop gain is 86.7 dB, and the unity gain bandwidth is 700 MHz. In S&H circuit, a 0.1 pF capacitance is connected to the non-inverting input terminal of the OA to store charge.

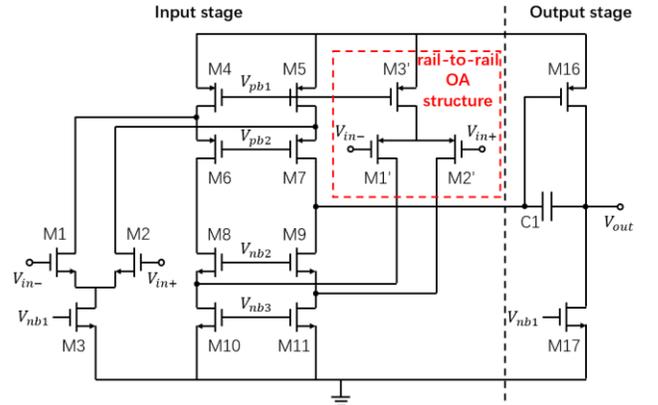

Fig. 5. OA design.

## IV. SIMULATION RESULT

To validate the AMC-based precoding architecture, the circuits were simulated in FreePDK, including the 45 nm RRAM addon process library [22]. The results of one simulation of the AMC architecture, including INV circuit, S&H module, and MVM circuit, are shown in Fig. 6. As illustrated in Fig. 6a, the INV circuit is powered by the applied input voltages $s$ at 10 ns. Within the next 10 ns, the INV circuit responds, and rapidly stabilizes to give the intermediate vector $y$. Such a fast response is attributed to the good condition of the matrix $Z$ in the massive MIMO model, according to the AMC theory that the convergence time is reciprocally determined by the minimal eigenvalue of the matrix [16]. Also, the optimized OA is important for the high speed of AMC circuits. At 20 ns, the transmission gates of the S&H module are turned on, the MVM circuit starts to work at the same time. Within the period 20 ns - 30 ns, the S&H module conveys the output vector $y$ of the INV circuit to the input terminals of the MVM circuit, the latter immediately establishing the stable output voltages $x$ at the crosspoint columns. The transient results of some representative nodes in the circuits are shown in Fig. 6b, which clearly demonstrates the fast response of the AMC circuits, completing the precoding computation within only 20 ns.

The precoded signal $x$ is transmitted through the massive MIMO channel according to Eq. (1). If the precoding is ideally precise, and the channel is free of noise, the received signal shall be precisely the ideal points in the constellation diagram. As the channel is always noisy, we tested the precoding efficiency under different SNR conditions. The introduction of channel noise cause deviations of the received signal in the diagram, which is reflected by a finite bit error rate (BER). In AMC precoding, as RRAM devices and AMC circuits show intrinsically non-ideal factors, they interact with the channel noise, contributing jointly to the BER performance. Fig. 7a and 7b show the signal transmission results under the SNR of 30 dB and 40 dB, respectively, both from 1000 random simulations. It can be observed that in the former case, the signals in the constellation diagram are still distinguishable, showing a BER value below $10^{-3}$. A higher SNR results in tighter distributions around the ideal positions, improving the BER to be $<10^{-4}$. We have systematically investigated the BER performance under different SNRs of the AMC-based precoding, in comparison with the results from conventional digital approach (Fig. 7c). Due to the AMC device/circuit non-idealities, there is a BER performance degradation in AMC precoding than the high-precision approach. Under relatively low SNR, such a degradation is alleviated, evidencing the tolerance of the model to non-idealities under noisy channels.

The most significant advantage of AMC for precoding lies in the extremely low computational complexity. Thanks to the massive spatial parallelism in the crosspoint RRAM array, both matrix inversion and multiplication are realized in one step. In this work, each matrix operation consumes merely 10 ns for a $16 \times 128$ MIMO precoding. By contrast, traditional digital algorithms generally exhibit high-order complexities. In Table I are listed three typical efficient matrix inversion algorithms for precoding, namely Neumann series [9], QR decomposition [6], and Gauss-Jordan elimination [7], which all show cubic complexities, although optimization schemes such as pipelining have been considered [10]. Digital algorithms might also be accelerated by using multiple parallel processors. For instance, in Ref. [10], 10 processors are implemented, and the resulting time of $16 \times 16$ matrix inversion is 1.96 µs. Matrix multiplication also shows a cubic complexity, although it can be more conveniently performed, e.g., through a distributed approach in the remote radio unit [10].

We have made an exhaustive estimation of the power consumption of the AMC precoding system, including the front-end DACs and the back-end ADCs. We refer to the reconfigurable SAR ADC designed specifically for analog computing with memory arrays [23]. We use 4-bit ADC and 2-bit DAC in the 16-QAM scenario. In addition, we considered the voltage follower OAs in the input circuit as well [24]. Finally, the power consumption proportion of different modules is shown in Fig. 8. Due to the large number and currents of MVM outputs, OAs in the MVM circuit account for half of the total power dissipation, which might be optimized by design strategies such as multiplexing and pipelining. By contrast, RRAM devices contribute a negligibly small fraction. The estimated power of the straightforwardly-a 16×128 MIMO configuration is 125 mW, which translates to 2.5 nJ energy for one vector precoding. By contrast, based on the result of a 8×128 MIMO system [25], the power of a digital processor for the 16×128 MIMO precoding is estimated to be 64 mW, which in turn suggests an energy cost of 125 nJ. Therefore, in this specific case, AMC precoder is 50× more energy efficient than the conventional digital approach.

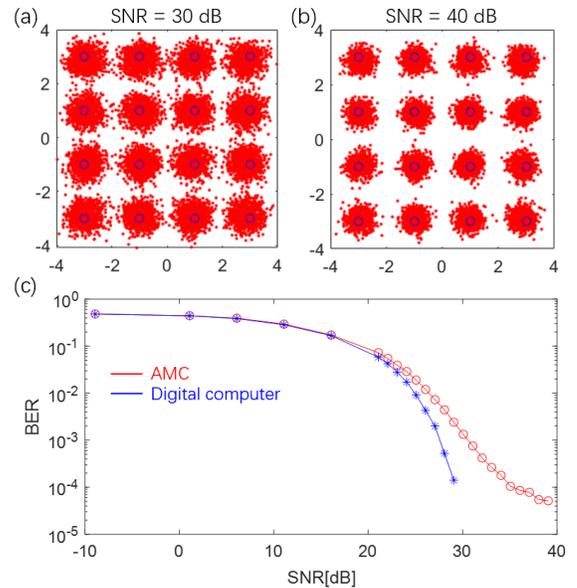

Fig. 7. Constellation diagrams of the received symbols (red points) after the precoding with AMC circuits, transmitted through noisy channels with SNR of (a) 30 dB and (b) 40 dB. The blue circles mark the ideal positions. (c) BER results of the symbol transmission precoded by AMC circuits and high-precision digital computer.

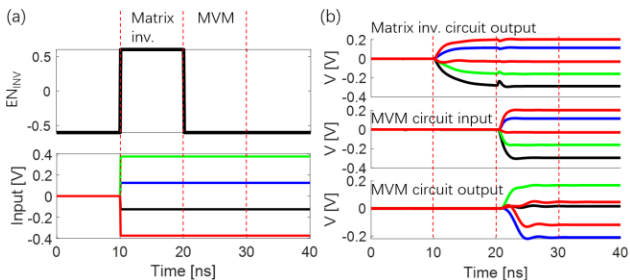

Fig. 6. (a) Wave forms of enable signal and representative input voltages. (b) Representative output and input voltage curves of AMC circuits.

TABLE I
COMPUTATIONAL COMPLEXITY COMPARISON

|  | Matrix inversion | Matrix multiplication |
|---|---|---|
| AMC | 1 | 1 |
| Nemann series | $K^3$ | $MK^2$ |
| QR decomposition | $3K^3 + 2K^2$ | $MK^2$ |
| Gauss-Jordan elimination | $K^3 + K^2$ | $MK^2$ |

## V. CONCLUSION

In this work, we applied the emerging RRAM-based AMC circuits to the linear precoding in massive MIMO for the first time. We demonstrated the high performance of this approach, including extremely low latency and high energy efficiency, by designing a new INV circuit, an analog cascading flow, as well as efficient OA circuits. Although the INV circuit is designed for ZF precoding in this work, it can be conveniently used for others popular scheme, *e.g.*, minimum mean square error (MMSE). We believe that RRAM-based AMC is highly suitable for MIMO computations, thanks to the moderate array size requirement, inherent error tolerance, and well-conditioned matrix structure. AMC represents a promising solution to the heavy matrix processing in massive MIMO systems, with an extension from this work to DFT/IDFT, detection, as well as channel estimation.

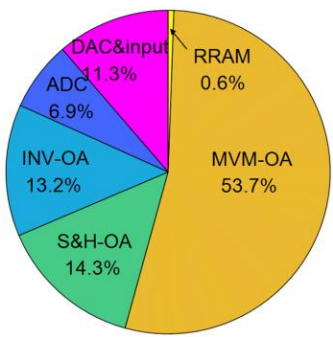

Fig. 8. Power consumption fractions of AMC precoding circuits.